\documentclass{article}

\usepackage[preprint]{neurips_2024}

\usepackage[utf8]{inputenc} %
\usepackage[T1]{fontenc}    %
\usepackage{hyperref}       %
\usepackage{url}            %
\usepackage{booktabs}       %
\usepackage{amsfonts}       %
\usepackage{nicefrac}       %
\usepackage{microtype}      %
\usepackage{xcolor}         %
\usepackage{graphicx}
\usepackage{natbib}
\usepackage{listings}    %

\definecolor{lightgray}{rgb}{0.95,0.95,0.95}

\lstdefinestyle{promptstyle}{
    basicstyle=\footnotesize\ttfamily,
    backgroundcolor=\color{lightgray},
    breaklines=true,
    breakatwhitespace=true,
    frame=single,
    framesep=5pt,
    numbers=none,
}

\usepackage{upquote}

\usepackage{booktabs}
\usepackage{siunitx}
\usepackage{multirow}
\usepackage{worldflags}
\usepackage{etoolbox}
\renewcommand{\bfseries}{\fontseries{b}\selectfont} %
\robustify\bfseries             %
\newrobustcmd{\B}{\bfseries}    %
\usepackage{tablefootnote}

\usepackage[textsize=tiny]{todonotes}

\title{
\centering
\raisebox{-0.25\height}{\includegraphics[width=0.07\textwidth]{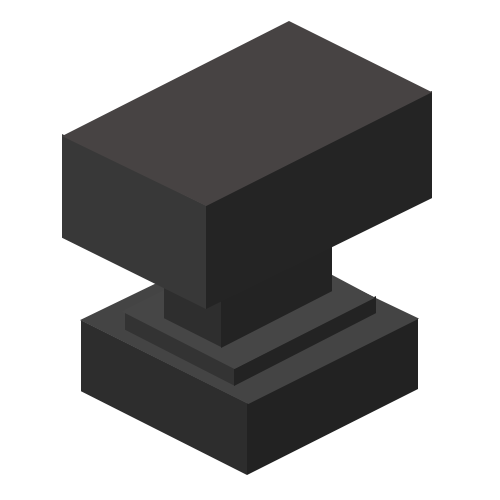}}
RepairBench: Leaderboard of Frontier Models for Program Repair
}

\author{%
André Silva and Martin Monperrus \\
KTH Royal Institute of Technology, Sweden \\
\texttt{\{andreans, monperrus\}@kth.se} \\
\\
\texttt{\url{https://repairbench.github.io/}}
\vspace{-0.7cm}
}

\begin{document}

\maketitle

\begin{abstract}
\vspace{-0.1cm}
AI-driven program repair uses AI models to repair buggy software by producing patches.
Rapid advancements in AI surely impact state-of-the-art performance of program repair. Yet, grasping this progress requires frequent and standardized evaluations.
We propose RepairBench, a novel leaderboard for AI-driven program repair.
The key characteristics of RepairBench are: 
1) it is execution-based: all patches are compiled and executed against a test suite,
2) it assesses frontier models in a frequent and standardized way.
RepairBench leverages two high-quality benchmarks, Defects4J and GitBug-Java, to evaluate frontier models against real-world program repair tasks.
We publicly release the evaluation framework of RepairBench. We will update the leaderboard as new frontier models are released.
\vspace{-0.2cm}
\end{abstract}

\section{Introduction}
\vspace{-0.3cm}

In recent years, AI-driven program repair \citep{zhang2023survey, zhang2024systematic} has emerged as a key application of AI in software engineering.
Program repair is the task of automatically fixing software bugs, and AI-driven repair uses AI models to generate bug-fixing patches.

Existing evaluation methodologies \citep{xu2022systematic, jiang2023impact} are inadequate for keeping pace with the rapid evolution of AI.
They fail to capture the longitudinal perspective required to track the progress of AI-driven program repair over new generations of AI models.
In this paper, we focus on frontier models, those state-of-the-art models that push the boundaries of AI capabilities.

We propose RepairBench, a novel leaderboard aimed at a frequent, sound, and standardized evaluation of frontier models for program repair.
RepairBench consistently evaluates frontier models on a high-quality set of program repair tasks.
RepairBench employs carefully curated benchmarks:
1) Defects4J v2 \citep{just2014defects4j}, a widely-adopted benchmark in the software engineering community, and
2) GitBug-Java \citep{silva2024gitbug}, a benchmark of recent bugs from 2023, that has been designed to address benchmark leakage.
A key design decision is that all bugs in RepairBench are real-world bugs coming from real-world programs. 
They also come with executable tests to verify the correctness of patches beyond syntactic match.

RepairBench carefully selects evaluation metrics to ensure a meaningful comparison across different models:
1) AST Match@1, which captures syntactic correctness w.r.t. the reference patch written by the human developer, and
2) Plausible@1, which captures correctness based on the execution of all test cases.
The latter is the default one used for ranking because it accounts for execution.

To sum up, our contributions are:
\vspace{-0.2cm}
\begin{itemize}
    \item \textbf{Leaderboard}: We publish RepairBench as a leaderboard on the web at \url{https://repairbench.github.io/}
    \item \textbf{Data}: We publicly share all prompts and patches at \url{https://github.com/ASSERT-KTH/repairbench}
    \item \textbf{Code}: We open-source the code to produce the leaderboard at:
 \url{https://github.com/ASSERT-KTH/elle-elle-aime}
\end{itemize}

\section{Methodology}

We devise the RepairBench methodology, a rigorous and standardized methodology to measure the performance of frontier models in program repair. This section outlines the key components of RepairBench, including the benchmarks, models, prompts, and the evaluation process.

\subsection{Models}

RepairBench exclusively focuses on frontier models.
Frontier models are models that, at the time of their release, stand out due to their performance across a wide-range of tasks when compared with the state-of-the-art.
Their capabilities are demonstrated in general-purpose \citep{hendrycksmeasuring, liang2022holistic, chiang2024chatbot} and code-specific tasks \citep{jain2024livecodebench}.
In other words, frontier models lie at the border of what AI models are currently capable of doing.

We select frontier models based on the following criteria:
1) they must demonstrate state-of-the-art capabilities (i.e., be frontier models) in other live evaluation systems (e.g., \citep{chiang2024chatbot, jain2024livecodebench},
2) they must be instruction-tuned, due to our prompt setup (see \autoref{sec:prompt}),
3) they must be available through an API that is accessible to the RepairBench team, and
4) the estimated cost to evaluate each bug in RepairBench must not exceed a given price (see \autoref{sec:costs}).

\subsection{Benchmarks}

RepairBench selects benchmarks per the following criteria:
1) being real-world programs (no toy programs, no competition programs),
2) being real-world bugs (no seeded or synthetic bugs),
3) including a variety of domains,
4) being executable, incl. at least one failing test case,
5) including a ground-truth patch written by a human developer, and
6) being well-engineered so that they can be integrated into the RepairBench framework with reasonable effort.

RepairBench V1 includes the only two benchmarks that meet all those criteria:

\textbf{Defects4J v2} \citep{just2014defects4j}, a widely-adopted benchmark in software engineering research, contains 835 real-world bugs from 17 open-source Java projects. We identify 484 single-function bugs which are utilized in RepairBench.

\textbf{GitBug-Java} \citep{silva2024gitbug}, is a benchmark of Java bugs from 2023, containing 199 real-world bugs from 55 open-source Java projects, from which we identify 90 single-function bugs utilized in RepairBench.

In the next update of the leaderboard, we plan to introduce SWE-Bench \citep{jimenezswe}.

\subsection{Prompts}
\label{sec:prompt}

RepairBench employs the same prompt setup for all models, to ensure consistency.
The prompt setup is zero-shot \citep{xia2022less}, targets single-function bugs (i.e., bugs whose reference patch alters a single function), and is not iterative \citep{zhang2024autocoderover, xia2024agentless} (i.e., only a single call to the model is made).
These choices are made for scoping reasons: RepairBench aims to provide a standardized evaluation of model capabilities, without accounting for additional approaches built on top of these models.

RepairBench's prompt template includes: 1) the buggy function, 2) the failing test cases' code, and 3) the failing test cases' error message (runtime information).
This set of ingredients captures the test-suite based program repair task \citep{parasaram2024fact}: the buggy function is the current program, and the failing test case/error provides the difference between current and expected behavior as defined by the developers.
All code snippets contain the original comments (e.g., inline comments, javadocs), and are surrounded by Markdown quotation marks.
Finally, the model is prompted to return the repaired function inside quotation marks.
\autoref{fig:prompt-example} shows an example prompt.

\begin{figure}[ht]
\centering
\begin{lstlisting}[style=promptstyle]
You are an automatic program repair tool. Your task is to fix the provided buggy code.

The following code contains a buggy function:
```java
    /**
     * Puts all values of this record into the given Map.
     *
     * @param map The Map to populate.
     * @return the given map.
     */
    <M extends Map<String, String>> M putIn(final M map) {
        for (final Entry<String, Integer> entry : mapping.entrySet()) {
            final int col = entry.getValue().intValue();
                map.put(entry.getKey(), values[col]);
        }
        return map;
    }

```

The code fails the following tests.

Test `org.apache.commons.csv.CSVRecordTest::testToMapWithShortRecord`:
```java
    @Test
    public void testToMapWithShortRecord() throws Exception {
       final CSVParser parser =  CSVParser.parse("a,b", CSVFormat.DEFAULT.withHeader("A", "B", "C"));
       final CSVRecord shortRec = parser.iterator().next();
       shortRec.toMap();
    }

```

Test `org.apache.commons.csv.CSVRecordTest::testToMapWithShortRecord` error:
```
java.lang.ArrayIndexOutOfBoundsException: 2
```


Please provide a fixed version of the buggy function, and only that function, inside a code block.
\end{lstlisting}

\caption{Prompt for bug Csv-6 of Defects4J. The test case and runtime information guide frontier models in generating patches.}
\label{fig:prompt-example}
\end{figure}

The answers generated by the models are expected to contain the fixed version of the buggy function inside quotation marks.
However, models are known to return additional natural language responses or explanations.
To retrieve the generated code with reasonable leeway for such text, we extract the first code block generated by the model using regular expressions.

\subsection{Costs}
\label{sec:costs}

Frontier models are typically expensive to evaluate due to both the energy cost to operate them and the provider's markup.
RepairBench is, for the most part, supported by the RepairBench team, who pay model providers for the patch generation jobs and who execute patches in local infrastructure.

To cap the amount of resources allocated to RepairBench, we define a maximum of 0.2 USD  per evaluated bug, or approx. \$115.1 for a total of 574 bugs. When a new frontier models is released, the RepairBench team estimates the cost to run RepairBench and proceeds only if the value is within the limit.
The cost to generate patches is calculated according to the pricing of each organization, or the pricing of third-party model providers in case of open-weights models.

RepairBench is open to sponsorship from model providers, in which case the cost threshold is not considered.

Cost is also important for program repair per se. Automated program repair fundamentally comptes with  the costs of human developers. RepairBench provides a cost-aware \citep{hidvegi2024cigar} view of program repair, and the trade-off between repair cost and repair effectiveness.

\subsection{Metrics}

The goal of program repair is to obtain a program that correctly fixes the bug without introducing any regression.
Thus, evaluating models for program repair involves evaluating the multiple dimensions of the patches generated by the models.

The patch should parse, compile, and type checks (depending on the target language).
Correctness is evaluated by running the repaired code against a set of test cases to ensure that the original issue is resolved without introducing new errors.  This is why we select benchmarks with reasonably good test suites.

RepairBench evaluates and ranks models according to two metrics. Both of them are meant to be maximized: the higher the metric, the stronger the model.

\textbf{Plausible@1}: the probability that the first generated patch passes all test cases. By running all test cases, we check if the original bug is resolved without new bugs being introduced. Note that this metric does not guarantee that the patch is functionally equivalent to the reference implementation since test suites typically do not cover the entire specification and input domains.
To compute the \textit{pass@k} metrics, we rely on Chen et al. \citep{chen2021evaluating}'s numerically stable and unbiased estimator, generating 10 non-deterministically sampled patches per bug with the provider's default settings and a temperature of $1.0$.

\textbf{AST Match@1}: the probability that the first generated patch has the same abstract syntax tree (AST) as the reference patch provided by the benchmark. Unlike \textit{Plausible @1}, \textit{AST-Match @1} is static and does not rely on the test suite. This metric is a strong indicator of correctness: if the ASTs are the same, it means that the model was able to produce the exact same patch as the human developer.

Note that the \textit{pass@k} metrics are more reliable than simply computing the total number of correctly fixed bugs: 1) generating patches is not deterministic, even when using deterministic sampling algorithms \citep{ouyang2023llm}, 2) models are usually deployed with non-deterministic sampling algorithms in practice.
\textit{pass@k} accounts for the non-determinism by representing the probability of generating a correct patch given a budget of $k$ generations.

\section{Results}

This section contains the RepairBench results, and is structured to be updated over time with new frontier models. We plan to update the benchmarks for at least 3 years.
\autoref{tab:leaderboard} shows the leaderboard status as of \today.

\begin{table}[ht]
\centering
\makebox[\textwidth][c]{%
\resizebox{1.3\textwidth}{!}{
\large
\begin{tabular}{@{}ll S[table-format=2.1, detect-weight=true] S[table-format=2.1, detect-weight=true] S[table-format=4.2, detect-weight=true] S[table-format=2.1, detect-weight=true] S[table-format=2.1, detect-weight=true] S[table-format=4.2, detect-weight=true] S[table-format=2.1, detect-weight=true] S[table-format=2.1, detect-weight=true] S[table-format=4.2, detect-weight=true] c@{}}
\toprule
\multirow{2}{*}{\textbf{Organization}} & \multirow{2}{*}{\textbf{Model}} & \multicolumn{3}{c}{Defects4J v2 (484 bugs)} & \multicolumn{3}{c}{GitBug-Java (90 bugs)} & \multicolumn{3}{c}{\textbf{Total (574 bugs)}} & \multirow{2}{*}{Ref.} \\
\cmidrule(lr){3-5} \cmidrule(lr){6-8} \cmidrule(l){9-11}
 & & {Plausible@1} & {AST Match@1} & {Cost (\$)} & {Plausible@1} & {AST Match@1} & {Cost (\$)} & {\textbf{Plausible@1}\textsuperscript{1}} & {\textbf{AST Match@1}} & {\textbf{Cost (\$)}} & \\
\midrule
\worldflag[width=0.3cm]{US} Anthropic & claude-3-5-sonnet-20240620 & \B 41.5\% & 12.3\% & \$57.91 & 26.1\% & 9.0\% & \$30.20 & \B 39.1\% & 11.7\% & \$88.11 & \citep{claude-35-sonnet} \\
\worldflag[width=0.3cm]{US} OpenAI & gpt-4o-2024-08-06 & 34.1\% & 8.4\% & \$20.74 & 18.8\% & 8.1\% & \$9.77 & 31.7\% & 8.3\% & \$30.51 & \citep{gpt4o} \\
\worldflag[width=0.3cm]{US} Google & gemini-1.5-pro-001 & 30.3\% & \B 13.0\% & \$44.95 & 16.7\% & 9.6\% & \$33.70 & 28.2\% & \B 12.5\% & \$78.65 & \citep{gemini-1.5} \\
\worldflag[width=0.3cm]{US} Meta & llama-3.1-405b-instruct & 28.9\% & 7.7\% & \$17.42 & 16.7\% & 7.3\% & \$11.86 & 27.0\% & 7.6\% & \$29.28 & \citep{llama3} \\
\worldflag[width=0.3cm]{CN} DeepSeek & deepseek-v2.5 & 26.6\% & 6.4\% & \$14.17 & 17.6\% & 7.3\% & \$5.55 & 25.1\% & 6.5\% & \$19.73 & \citep{deepseekv2} \\
\worldflag[width=0.3cm]{CN} Alibaba Cloud & qwen-2.5-72b-instruct & 25.5\% & 6.7\% & \$2.46 & 17.3\% & 5.9\% & \$2.28 & 24.2\% & 6.6\% & \$4.74 & \citep{qwen2.5} \\
\worldflag[width=0.3cm]{EU} Mistral & mistral-large-2407 & 24.5\% & 6.6\% & \$27.17 & 15.2\% & 6.6\% & \$20.53 & 23.0\% & 6.6\% & \$47.70 & \citep{mistral-large} \\
\worldflag[width=0.3cm]{US} OpenAI\textsuperscript{2} & o1-preview-2024-09-12\textsuperscript{2} & \multicolumn{1}{c}{---} & \multicolumn{1}{c}{---} & \multicolumn{1}{c}{---} & \B 32.3\% & \B 12.1\% & \$325.71 & \multicolumn{1}{c}{---} & \multicolumn{1}{c}{---} & \multicolumn{1}{c}{---} & \citep{o1} \\
\bottomrule
\addlinespace
\multicolumn{12}{l}{\textsuperscript{1}Models are sorted by the total Plausible@1 score.} \\
\multicolumn{12}{l}{\textsuperscript{2}Only partial results available right now due to cost reasons.} \\
\addlinespace
\end{tabular}
}
}
\caption{Leaderboard of Frontier Models for Program Repair as of \today}
\label{tab:leaderboard}
\end{table}

The leaderboard highlights a clear dominance of Anthropic's \textit{claude-3-5-sonnet-20240620}, which achieves the highest overall Plausible@1 score ($39.1\%$). This means that this model captures the most of the expected behavior specified in one shot prompt, coupled together with perfect mastering of the syntax of the programming language.

OpenAI's \textit{gpt-4o-2024-08-06} and Google's \textit{gemini-1.5-pro-001} achieve the second and third best scores, respectively.
\textit{gemini-1.5-pro-001} is the best model according to AST Match@1 ($12.5\%$).

OpenAI's \textit{o1-preview-2024-09-12} results are currently incomplete due to its high cost. Yet, we note that it achieves the best score on GitBug-Java with a $32.3\%$ Plausible@1 score (as opposed to $26.1\%$ for \textit{claude-3-5-sonnet-20240620}).

\begin{figure}[h!]
    \centering
    \includegraphics[width=0.75\linewidth]{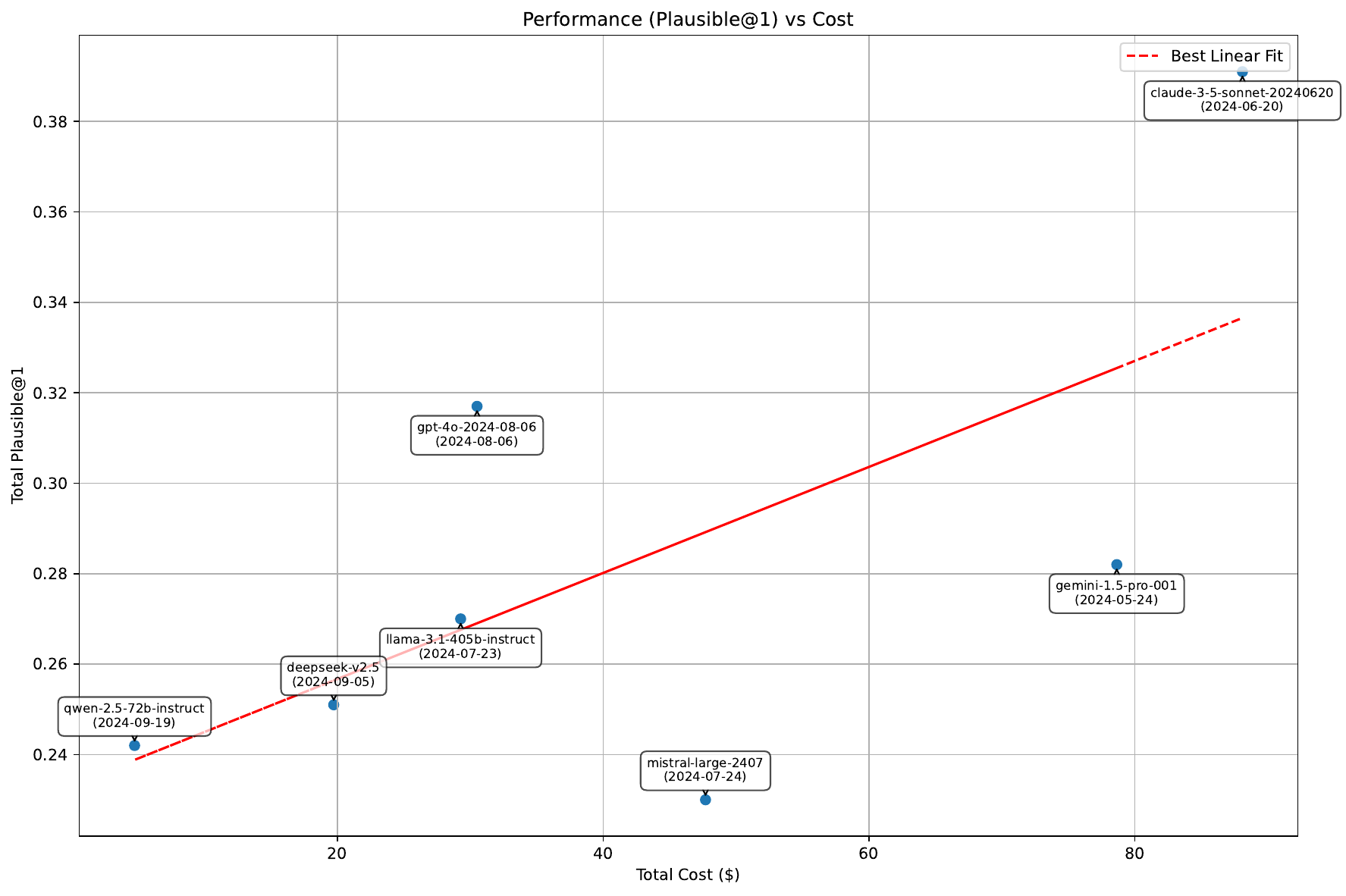}
    \caption{Performance (Plausible@1) in function of the total cost (in USD). The most performant models are also the most expensive ones.}
    \label{fig:performance_cost}
\end{figure}

\autoref{fig:performance_cost} plots the performance (Plausible@1) as function of the total cost to run RepairBench on each model.
The most expensive frontier models are also the most performant.
Anthropic's \textit{claude-3-5-sonnet-20240620} costs a total of \$88.11 for a score of 39.1\%.
Alibaba Cloud's \textit{qwen-2.5-72b-instruct} model, the cheapest model (4.74\$, approx. 20x less than \textit{claude-3-5-sonnet-20240620}) in RepairBench, achieves a score (24.2\%) comparable with DeepSeek's \textit{deepseek-v2.5} (25.1\%) and better than Mistral's \textit{mistral-large-2407} (23.0\%).

\section{Discussion}

A critical concern when evaluating AI models is the issue of benchmark leakage \citep{dong2024generalization, matton2024leakage}.
Benchmark leakage occurs when models are exposed to test data during pre-training or post-training.
Benchmark leakage, aka contamination, can lead to inflated performance results, giving a misleading picture of the actual ability to generalize and solve novel problems.
This concern is particularly problematic for frontier models due to their training on huge amounts of data.

We believe there is some benchmark leakage for Defects4j, but the actual extent is unknown. However, the low overall performance (approx. 30-40\%) shows that the benchmark is not at all perfectly memorized.

In RepairBench, we do mitigate benchmark leakage with the inclusion of GitBug-Java \citep{silva2024gitbug}, a newly constructed benchmark with only recent bugs, from 2023 onwards.
Moreover, RepairBench prioritizes an execution-based metric (Plausible@1) over purely static evaluations (AST Match@1): this helps assess model capabilities beyond superficial memorization of code.

\section{Related Work}

\subsection{General-Purpose Benchmarks}

The evaluation of AI models typically relies on general-purpose benchmarks that assess performance across diverse domains.
Among these, MMLU \citep{hendrycksmeasuring} stands out as a comprehensive benchmark, encompassing problems from a wide array of academic disciplines.
HellaSwag \citep{zellers2019hellaswag} focuses on testing models' commonsense reasoning, while benchmarks like GSM8K \citep{cobbe2021training} and MATH \citep{hendrycks2measuring} are designed to evaluate models' mathematical problem-solving capabilities.

In parallel, several live evaluation platforms have emerged to continuously measure model performance.
HEML \citep{liang2022holistic} aims to provide a comprehensive evaluation of models across a range of tasks.
Other platforms, such as Vellum\footnote{\url{https://www.vellum.ai/llm-leaderboard}}, Open LLM Leaderboard\footnote{\url{https://huggingface.co/spaces/open-llm-leaderboard/open_llm_leaderboard}}, and KLU.ai's leaderboard\footnote{\url{https://klu.ai/llm-leaderboard}} also provide live updates of model performances.
Notably, ChatBotArena \citep{chiang2024chatbot} maintains real-time leaderboards based on battles between models and a crowd-sourced evaluation methodology.

While these benchmarks cover a broad range of capabilities and reasoning tasks, none of them address the specificity of the program repair task. 

\subsection{Code Benchmarks}

Code-related tasks, such as code generation and repair, require specialized benchmarks.
Being code, execution is a unique characteristic of the output and
we claim that execution-based code benchmarks is crucial \citep{khan2024xcodeeval}.

One of the most widely-used execution-based benchmarks in this area is HumanEval \citep{chen2021evaluating}, which evaluates the ability of models to generate Python code for simple algorithmic problems.
Although HumanEval has been an important tool for measuring the effectiveness of models in code generation, it is now exhausted, as frontier models achieve near-perfect scores.
Also, it is not a program repair task.

Program repair benchmarks \citep{le2015manybugs} provide a suitable testing ground for AI-driven program repair.
Several program repair benchmarks have been proposed across languages and domains \citep{csuvik2022fixjs}, \citep{gyimesi2019bugsjs}.
Some program repair benchmarks are exclusively static, without test cases available for execution \citep{avulaminecraft}.
RepairBench only focuses on executable benchmarks.

Other benchmarks, despite including test cases, are not fully reproducible due to missing third-party dependencies and other low level problems \citep{madeiral2019bears, saha2018bugs}.
RepairBench only focuses on reproducible benchmarks \citep{zhu2023reproducibility}.

\subsection{Code Leaderboards}

Beyond sporadic evaluations, live evaluation platforms for code have been proposed.

Aider's leaderboard\footnote{\url{https://aider.chat/docs/leaderboards/}} evaluates LLMs on their capability to write code according to a given instruction.
In contrast, RepairBench focuses exclusively on program repair, which involves fixing real-world bugs in existing codebases.

LiveCodeBench \citep{jain2024livecodebench} offers a continuous evaluation of LLMs on a variety of code-related tasks, including self-repair \citep{fan2023automated}, where the model is assessed based on its ability to fix code it has previously generated.
While LiveCodeBench focuses on artificial tasks extracted from code competitions, RepairBench only evaluates models with real-world repair tasks that human developers have encountered during software development.
Finally, Shariffdeen et al. \citep{shariffdeen2023program} held a competition of program repair approaches, but do not include frontier models.

\section{Conclusion}

RepairBench introduces a standardized, execution-based evaluation framework for assessing frontier models in AI-driven program repair.
RepairBench relies on real-world bug benchmarks and focuses on execution for evaluating patches.
As new frontier models will be released, RepairBench's leaderboard will provide insights into the longitudinal evolution of AI-driven program repair.

\section{Acknowledgments}

This work was partially supported by the Wallenberg AI, Autonomous Systems and Software Program (WASP) funded by the Knut and Alice Wallenberg Foundation.
The computations/data handling were enabled by the supercomputing resource Berzelius-2023-175 provided by National Supercomputer Centre at Linköping University and the Knut and Alice Wallenberg foundation.

\bibliographystyle{plainnat}
\bibliography{references}

\end{document}